\documentclass{article}
\usepackage{graphicx}
\pdfoutput=1
\graphicspath{}
\begin{document}
\centerline{\bf Kuiperian Objects and Wandering Cosmic Objects.}
\centerline{by}
\centerline{Gilles Couture}
\centerline{D\'epartement des Sciences de la Terre et de l'Atmosph\`ere}
\centerline{Universit\'e du Qu\'ebec \`a Montr\'eal}
\centerline{CP 8888, Succ. Centre-Ville}
\centerline{Montr\'eal, QC, Canada, H3C3P8}
\centerline{(couture.gilles@uqam.ca)}
\begin{abstract}
We study the effects of an encounter between a wandering cosmic object 
(WCO) of 0.1 solar mass and some Kuiperian Objects (KO). 
First, we let the WCO cross the out-skirt of our Kuiper belt. 
Such encounters can produce 
two types of solar objects: Eccentric Kuiper Objects of type I (EKO-I) whose 
perihelion is comparable to, but always smaller than the aphelion of the initial 
KO and Eccentric Kuiper Objects of type II (EKO-II) whose perihelion 
can be as small as a few AU. EKO-I tend to have a fairly large range of 
eccentricities, but EKO-II tend to have very large eccentricities. Both 
tend to be produced in clusters similar to those observed in Extreme 
TransNeptunian Objects (ETNO). 

When a WCO crosses the path of an EKO-I, it will produce two main  
classes of objects: Far Kuiper Objects (FKO) of types I and II. 
The Sednitos discovered in the past years fit the FKO-I class 
with their large major axis and fairly large eccentricity, while the FKO-II class is 
different with its large major axis but smaller 
eccentricity and opposite spinning direction. When 
a WCO encounters an EKO-II, the latter can remain in the same class, spinning in either 
direction, it can end up in the EKO-I class also spinning in either 
direction, but it can also 
be sent onto orbits with extremely long semi-major axis,  
relatively small eccentricity where both spins are allowed
This FKO-III class could be likened to Lower Oort Cloud 
Objects as their major axis is a fair fraction of a light-year.

These results lead us to consider the possibility that the Kuiper Belt was once substantially 
larger than it is now, perhaps 90 AU. 
We find some evidence of this 
scenario in current astronomical data. 
\end{abstract}
\vfil\vfil\eject

\noindent{\bf {Introduction}\hfil\hfil}
\hfil\hfil\break\noindent
The Kuiper Belt is a very complicated dynamical system whose current configuration likely  
carries traces of the genesis of the solar system. Its different populations 
[Gladman et al., 2008] have been influenced by the early solar environment and it has been shown 
[Ida {\it et al.}2000] that their orbital distributions present evidence of interactions with neighbour 
stars similar to the Sun at this early epoch. 
Several models of genesis have been tested with ever increasing 
numbers of elements [Gladman et al., 2001, Morbidelli et al. 2007,  Morbidelli \& Nesvorny, 2019].
Interestingly, among the different populations, it seems binary systems are in fact fairly common and 
play an important role. 
[Thirouin, Sheppard \& Noll, 2017, Thirouin \& Sheppard, 2018, Fraser et al., 2017]
Once a population is created, its evolution will be subjected 
to its environment. Complex resonances coupling elements of the Kuiper belt to our giant 
planets have been observed over the past decades [Bannister et al. 2016; Holman et al. 2017]
indicating clearly that they have played a role in shaping the belt.
Multiple simulations [Levison et al. 2007, Brasser \& Morbidelli, 2013] over the years have 
shown that the fate of planetesimals can be dramatic 
in an open clusters where stars are formed: they can be exchanged between stars or even 
ejected altogether [Hands et al., 2019]. The early environment of the sun has implications as 
far away as the inner Oort cloud system [Brasser et al., 2011]

A recent global survey  [Bannister {\it et al}, 2018] has provided a wealth of information 
that will not only constrain these models but also constrain the populations of objects that 
lay beyond the Kuiper belt as they would  affect the Kuiperian elements [Kaib et al. 2019].
Interestingly, the population of craters observed recently on Pluto and Charron 
by New Horizon [Singer, K.N., 2019] seems to indicate a deficit in the population of small 
objects in the Kuiper belt.

It has been observed that the Kuiper belt has a rather sharp edge at 
about 50AU.[Allen et al., 2001] The origin of this edge is still not well understood 
but some stellar flyby could have played a role in its creation [Kenyon \& Bromley, 2004]
Beyond the edge, exists a population of elements 
whose eccentricities are far larger than those of the main belt. These elements are still 
under Neptune's influence when at their perihelion and 
interactions between them could be at play in producing their very different orbits.
Different models were proposed to try to understand the production of these objects 
[Morbidelli et al. 2004; Trujillo \& Brown, 2001]

Beyond these objects lay the elements 
that are not affected by Neptune, their perihelion being over 60 AU 
[Gomes et al. 2008, Brasser \& Schwamb, 2015]. 
The first one to be discovered was Sedna [Brown, Trujillo, \& Rabinowitz,2004]. 
A previous object (200CR105) had already  attracted a fair amount of attention because of its 
large aphelion. Gladman et al., (2002) had shown that it could not be a regular member of the 
scattered disk that had its orbit modified by direct gravitational scattering off any giant planet. 
Morbidelli and Levison (2004) proposed some  
scenarios that involved different massive objects at different epochs of the solar system or the 
passage of nearby stars in order to explain its orbital parameters and applied the same scenarios 
to 2003VB12 (Sedna). For a little while Sedna was alone in its class but the discovery of 
2012VP113 brought a second member. [Trujillo \& Sheppard, 2014]. In the course of these studies, it was 
noticed that the distribution of these objects in space was unusual: several had a tendency to aggregate in  
a relatively small portion of the sky. This was explained by one [Trujillo \& Sheppard, 2014] or more 
[de la Fuenta Marcos \& de la Fuenta Marcos, 2014] heavier trans-Plutonian planets. Numerical simulations were 
later performed and the Planet Nine (NP) hypothesis was proposed [Batygin \& Brown, 2016].
In order to be able to shepherd these objects in a small part of the solar system, this massive object 
would need a mass of about 16 Earth masses, and an orbit of 
comparable eccentricity (about 0.6) with an aphelion of about 120 AU. 
Current planetary models allow an estimate of its structure [Linder \& Mordasini, 2016].

Shortly after NP was suggested, a fairly large section 
of its orbit was ruled out through a careful analysis of 
Cassini data; although very far from us, NP would have a small effect 
on Saturn [Fienga, A., et al., 2016]. 
A mean motion resonance mechanism between NP and 
other Sednitos was suggested as the source of the clustering 
of the Sednitos' orbits [Batygin \& Brown, 2016] 
and more detailed studies followed [Malhotra et al., 2016;
Millholland \& Laughlin, 2017].
The effect of NP on the mass distribution within the 
Kuiper belt has been calculated [Lawler et al., 2016] and while the distributions 
with and without NP are qualitatively different, it could be 
very difficult to distinguish them observationally.

The recent discovery of 2015TG387 [Sheppard, Trujillo, Tholen \& Kaib, 2019] brought a third 
member to the group of Neptune-free objects. We will also include 2013SY99 in the same 
category as Sedna because of its large perihelion [Barrister et al., 2017] It is still difficult 
to understand how these objects got to such large orbits. Were they formed at such distances 
from the sun or were they formed much closer and then moved far away ? 
If they were moved far away, what process was involved [Stern, 2004]? 
Could this process have moved many more of these objects ? 
The stellar flyby that could 
have shaped the edge of the Belt could also have sent objects on these exotic orbits 
[Kenyon \& Bromley, 2004]. One potential problem with a stellar flyby is that if 
a star came close to our solar system in the past, it is very likely that we would have 
discovered it by now as a star with very little angular motion. Unless the process took 
place during the genesis of the solar system in which case the companion 
stars of the sun might be very, very far away.

The hypothesis of a ninth planet is still debated and will remain so until 
NP is discovered. It has been argued that the orbits of some Sednitos have been wrongly 
estimated, thereby weakening the reason to invoke NP at all.[Shankman et al., 2017]; 
although recent results tend to agree with the 
presence of a ninth planet [de la Fuente Marcos et al. 2017] There is also the scenario 
where all Sednitos were captured by the sun during a close encounter with a star and 
its own planets [Jilkova et al., 2015].  Batygin \& Morbidelli (2017) have studied the 
possibility that some distant KO be trapped in some resonances with NP while 
Sheppard \& Trujillo (2016) note that correlations among several extreme eccentric objects 
is further evidence of a distant massive planet. The discovery of 2015BP519 
[Becker et al., 2018] was seen as further proof of its existence 
because it can produce objects with orbits highly inclined 
with respect to the global orbital plane. Recent work however  
[de la Fuente Marcos et al., 2018b] warns of some caution when using 2015BP519 
as further proof of the existence of PN because it appears as a statistical outlier.

Recently, a cosmic messenger has been discovered: Oumuamua.[Bacci et al., 2017, 
Meech et al. 2017] This object is also very special because it is believed 
[Bannister et al, 2017b; Knight et al. 2017; Meech et al. 2017b] to be the first interstellar 
planetesimal that we see. Its discovery is taken as a clue that such 
objects must be fairly common (at least not so rare) in interstellar 
space. [Pfalzer \& Bannister., 2019]

Five years ago [Scholz, 2014], a faint binary system consisting of a small red dwarf 
and a smaller brown dwarf with masses of about 8\% and 6\% of a solar mass 
was discovered and named WISE J072003.20-084651.2.  
About two years later, careful measurements [Mamajek et al. 2017] 
showed a very small angular motion that indicates that this star 
was at some point in the past very close to our solar system. It is estimated that 
Sholz's star is now about 20 LY from us and from its speed of about 
100 km/sec, it came within about 0.75  LY from our solar system 
about 70000 years ago. Albeit small, this system  
seems to have had some effect on the radiants observed 
in our extended solar system [de la Fuente Marcos, de la Fuente Marcos, Aarseth, 2018]. 
The origin of this particular WS is still not well understood, but there are some ways to 
produce a wandering stars: a double star system that 
gets too close to a cluster within a galaxy and is broken up by 
tidal forces, a double star system that gets too close to a black hole at the center of a galaxy and 
looses a member to the black hole. One could 
also think of a star being ejected from its home galaxy during an encounter with 
another galaxy, and, not being able to catch up with the visiting galaxy ends up as a wandering star.

Similar to the observation of Oumuamua, one could take the observation of Schulz's star 
as a clue that such lone, very light stars might not be so rare in the cosmos.

In this paper, we want to quantify the effects of the passage of a wandering 
cosmic object (WCO) on objects that are on the outskirt of our Kuiper belt. 
We will chose the WCO as rather light, 0.1 solar mass, so 
as to be barely a star and almost invisible. If a star would have passed close 
to our solar system, we would have seen it already. Therefore, in order for this 
object to not have been detected, it must be very dim, while being relatively
massive in order to have an effect on the outer solar system when it passes by. 
A light star seems the most appropriate object. Although an object of about 
one solar mass would also be possible as a light white dwarf, a light neutron star 
or a very small black hole, 
the light star seems the less exotic scenario. One could also think of a 
collection of several small objects (each one with a mass of a few percent of 
a solar mass) but this scenario seems less likely as the cluster might have been 
pulled apart on its journey, similar to the Schoemaker-Levy comet of 1994. Regarding its 
speed, we chose 50 km/sec. This value is less than the estimated speed of Scholtz's star
but higher than that of Oumuamua which is estimated at 25 km/sec.  
We also do not want to perturb too much the objects of our Kuiper belt and our 
way of achieving this is to assume that the wandering cosmic object will  
barely meet the out-skirt of the belt. This way, it could potentially encounter
several objects that it could send to another orbit and also eject several 
objects in outer space, while leaving most of the Kuiper belt unaffected.  
Such an object has no effect on the inner solar system and very little effect
on the outer planets.

In a first section, we will describe the procedure 
that we used to produce the three body system. Then, we will briefly describe the 
3 sections of our calculations: the interaction of the CWO with the outs-kirt of 
the Kuiper Belt and then the interaction of the CWO with two objects produced by 
the first interaction. The figures that we present in the paper are the regular 
$x-y$ plane with units in $m$. They are complemented 
by tables where we give precise values of the orbital parameters; these are 
tables 2, 3, and 4. In the first table, we present the orbital parameters of some 
Trans-Neptunian and Extreme-Trans-Neptunian objects; 
we categorize them as Class-I, II, and III. This is not the regular classification 
used for these objects [Sheppard et al. 2019], 
but it is relevant for the orbits produced in the 
process studied here. 
We do not include 2018VG18 in our list since 
its orbital parameters are not very well known yet. Note also that we include 
2013SY99 in the Class-III objects (as opposed to Class-I) because of its 
perihelion of 50 AU, which is very close to the outer edge of the Kuiper Belt.

\hfil\hfil\break
\noindent{\bf {Calculations}\hfil\hfil}
\hfil\hfil\break\noindent
The calculation was performed as follows. We have an absolute frame of 
reference where we put the Sun, Kuiperian object, and CWO. Our input 
parameters are the masses, initial coordinates, and velocities of the 
3 objects. The initial coordinates of the objects are $(0,0)$ for the 
sun, $(q,0)$ for the Kuiperian object and $(x_0,y_0)$ for the CWO. The 
sun is initially at rest. As for the Kuiperian object, once we decide the 
value of $q$ and $\varepsilon$ (eccentricity) its speed is calculated at 
$t=0$ . As for the CWO, we specify its initial position and velocity 
(in most cases, its velocity will be 50 km/sec) 
depending at what point we want it to meet the Kuiperian object. 
Once these initial parameters are defined, we use a few analytical time steps 
to start the motion properly and then we let the system evolve 
using $\vec F = m\vec a$ for a time step that is as small as possible. 
Depending on the orbits, our time step could 
be as large as 200-300 seconds and as small as 10 seconds. Once the 
calculation is done in the absolute frame, we simply bring back all 
coordinates and speeds in the frame of the sun, since this is our reference 
frame: in the figures presented here, the sun is always at the origin.  
When we use the absolute frame, we see that the sun moves and the path of the 
CWO is not a straight line. 

\hfil\hfil\break\noindent
\noindent{\bf {Kuiper belt}\hfil\hfil}
\hfil\hfil\break\noindent 
In this calculation, we neglect the mass of the Kuiper belt and consider only the 
interactions between a member of the belt, the sun and the CWO. 
At $10^{10}$ kg for our KO, one
could argue that this mass is very small and some objects could have
a much larger mass. We have verified that we can push this mass up
to $10^{20}$ without affecting our results; as long as the mass of the
KO remains negligible in front of a solar mass, our results hold.

We take the Kuiper belt as a disk with inner radius $q_K$ and 
outer radius $Q_K$ where the sun is at the center, the origin of our axes. 
The Kuiperian objects move along ellipses of small 
eccentricities within the disk and 
we make them rotate counter-clockwise always starting on the $x-axis$. 
Now, let a straight line cut this disk vertically but 
not too far from the edge; the minimal distance from this line to the center is a 
little smaller than $Q_K$. The entrance and exit points are A and B, respectively. 
Assume the CWO follows this straight line upward and comes from the bottom. If it crosses 
the disk left of the center, it will encounter the Kuiperian objects in a head-on 
collision with points A and B while if it crosses the disk on the right of the 
center, we will have a rear-end collision with entrance and exit points as C and D.

In order to see clearly the dynamics of the process, we will allow our 
CWO to cut the Kuiper Belt at a fair distance from the edge. Afterwards, we will see 
what happens when it is allowed to meet the belt at a grazing angle. 

\hfil\hfil\break
\noindent{\bf {Head-on collision}\hfil\hfil}
\hfil\hfil\break\noindent
\noindent{\bf {Entrance point, A}\hfil\hfil}
\hfil\hfil\break\noindent
We will use a KO with 
$q = 7.6\times 10^{12}m$ and and eccentricity of 0.043, which leads to 
$Q = 8.28296\times 10^{12}m$. We will let the WCO start its journey at initial 
coordinates $(-7.8\times 10^{12}, y_0)$ and velocity $(0,50000)$ m/sec. Then we let 
the system evolve. The CWO is not deflected much by the sun and comes closest at 
a distance of about $7.742\times 10^{12}$ m. We simply chose the initial distance 
$y_0$ depending where we want it to encounter the Kuiperian object. On Figure 1-A and 
Table 2, we 
show several orbits such an encounter can produce. The initial orbit is the ellipse 
in dotted black at the center. Each orbit is the result of a particular encounter: 
the order of decreasing-increasing minimal distance between the KO and the CWO is 
red-green-dark/blue(dotted)-magenta-light/blue-black. As the encounter gets closer, 
the orbit is pulled down from the initial orbit to the red orbit. Then, it becomes a 
highly eccentric orbit (green) as the minimal distance decreases before it produces 
less eccentric orbits as the minimal distance increases again and finally 
has smaller effect on the orbit as the minimal distance is fairly large.  
It is important to note that 
all these orbits are rotating counter-clockwise, (as the initial orbit) are closed, 
some extend very far, 
as can be seen in Table 2, where the angle $\theta_x$ is defined as the angle of $Q$ (taken as 
vector from the sun to the aphelion) with respect to the positive $x-axis$. In these encounters, 
the highly eccentric orbits (green dots) 
are produced not by sending the KO toward the sun, but by sending it first in outer space.   
Also important and not shown here is that a whole class of open orbits are sent first towards 
the sun before being ejected; some get very close to the sun, and some are even ejected on a 
clockwise open orbit. 
These orbits happen between the green and dark-blue orbits 
on this figure. 

We wee that the window to produce highly eccentric orbits (EKO-II) is rather small 
compared to that 
for producing less eccentric orbits (EKO-I). We also see that the orbits have a 
tendency to cluster, both 
the EKO-I and the EKO-II, albeit at different angles. One must note also that it is impossible to 
produce 
an object such that $q> Q_{initial}^{kuiper}$.  

\hfil\hfil\break
\noindent{\bf {Exit point, B}\hfil\hfil}
\hfil\hfil\break
On its way to the exit point, the effects of the CWO are similar to that of the black 
curve on Figure 1-A. Of course, this depends on the entrance and exit points: as A and 
B are more separated, there will be a window where the effect of the CWO will be very 
small, as it is farther and farther away from the KO on its journey from the entrance 
to the exit point.  

On Figure 1-B and Table 2, 
we use the same colour code for the decrease-increase of the impact parameter between 
the CWO and the KO as for the entrance point. The initial orbit is the black-dotted 
ellipse in the center. As the WCO gets closer to the KO, it will drag the orbit and 
bring it to the red orbit. As the impact parameter decreases, the orbits will turn into 
very elongated orbits with small $q$ (green) However, contrary to the entrance collision, 
the KO is now sent first toward the sun. As the impact parameter increases, the collision 
produces the not so eccentric orbits with fairly large values of $Q$ (dark-blue and magenta) 
before leaving the system and producing orbits light blue and black. Again, there is a whole
class of orbits that are sent toward the sun, some end up very close to the sun and can 
even change rotation. These orbits happen between the highly elliptical orbits (green) and 
the not so elliptical ones (dark blue) Again, all these orbits were ejected from the solar system.

As for the entrance point, we see that the window for producing EKO-II  
is rather small and it is impossible to produce an object such that $q> Q_{initial}$. Again, 
both EKO-I and EKO-II have a tendency to cluster. In this configuration, the angles between 
the two EKO-I clusters (entrance and exit) is about 170 degrees and that between 
the two EKO-II clusters is about 120 degrees. Note also that the two clusters 
are or either side of the incoming trajectory of the CWO.  

In general, the CWO will get closer to the KO in an encounter that will produce an EKO-II than 
it will in one that will produce an EKO-I. If the CWO gets too close, the KO will be simply ejected 
from the solar system.
 
\hfil\hfil\break
\noindent{\bf {Rear-end collision}\hfil\hfil}
\hfil\hfil\break
\noindent{\bf {Entrance point, C}\hfil\hfil}
\hfil\hfil\break\noindent
We use the same KO, ({\it ie} $q = 7.6\times 10^{12}m$ and $Q = 8.283\times 10^{12}m$) 
but since it is now the rear-end collision scenario, we will 
use the following initial coordinates for the WCO 
$(7.1\times 10^{12}, y_0)$ and velocity $(0,50000)$ m/sec, and we vary $y_0$ to 
get the WCO to meet the KO at the desired point. 
The WCO is not affected very much by the sun and its closest distance to the sun will be 
$7.042\times 10^{12}$ m.

Figure 1-C and Table 2 show the different orbits such an encounter can produce. All these orbits are
closed and in the same rotation as the initial object, counter-clockwise. The initial orbit 
is the dotted black ellipse in the center of the figure. We use the same colour code as before 
to show the decrease-increase in the impact parameter between the CWO and the KO. When the 
impact parameter is large, not much happens and as the impact parameter decreases, the orbit 
is perturbed and produces the large orbits represented in red. As the impact parameter decreases 
the KO is ejected in outer space before being deflected toward the sun and again ejected from the 
solar system. Then, the deflection toward the sun produces closed but very elongated orbits 
in dotted green, blue and magenta ellipses. As the impact parameter continues to increase, the 
collision produces less eccentric orbits (light blue and black) until the orbit is perturbed 
very little. It is important to note here that the highly eccentric orbits are produced by 
sending the KO first toward the sun. There is a window of impact parameters that will send 
the KO out of the solar system either by sending them first toward the sun or directly 
into outer space.

\hfil\hfil\break
\noindent{\bf {Exit point, D}\hfil\hfil\hfil}
\hfil\hfil\break\noindent
Figure 1-D and Table 2 show the different orbits that we can obtain when the WCO exits crosses 
the orbit of a KO as it moves out. The structure is a little simpler here: as the impact 
parameter decreases, the orbit is 
pulled down and amplified (red) until it reaches very large aphelion (green). 
Then, the impact parameter decreases some more and the KO is sent toward the sun before 
being ejected from the solar system. As the impact parameter increases a little, the 
KO is sent to outer space on long, highly eccentric orbits (dotted blue and magenta) 
Finally, as the impact parameter increases, the orbits settle on less eccentric 
orbits that will be closer and closer to the original orbits. In this figure, all 
orbits have the same rotation as the original orbit (counter-clockwise). One must note 
here that it this collision process, it has been impossible to send the original KO 
toward the sun and obtain a closed orbit. All orbits that send the KO toward the sun 
eject the KO from the solar system. One must also note that the range of values for 
$q$ to produce long, eccentric orbits is rather small (dotted blue and magenta)
Again, such collisions cannot produce a perturbed orbit where $q > Q_K$.

We also note clustering of the EKO-I and EKO-II. The clustering is a little different depending 
whether we have a head-on encounter or a rear-end encounter. The EKO-I will be more back-to-back 
in a head-on encounter than in a rear-end encounter: about 170 degrees vs about 140 degrees in our 
case. The EKO-II will be less back to back in a head-on collision than in a rear-end collision: 
about 120 degrees vs about 140 degrees in our case.

\hfil\hfil\break
\noindent{\bf {Discussion}\hfil\hfil}
\hfil\hfil\break\noindent
We see that both types of orbits (EKO-I and EKO-II or Class-I and Class-II) are produced 
in the collision process. There is however a difference in their respective windows: the 
window to produce EKO-I is larger than that to produce EKO-II. Furthermore, there is also 
a difference in the range of values allowed for $q$ in an EKO-II orbit depending whether 
we have a head on collision or a rear end collision: the range is smaller when the 
perturbed orbit is produced at the exit point in a rear end collision.


Regarding the clustering, we note that EKO-I are produced on the opposite side of the 
trajectory of the CWO with respect to the sun while the EKO-II are produced on the same 
side. This clearly means that the passage of a single CWO through the edge of the 
Kuiper belt would produce several objects of EKO-I and EKO-II classes that would not be 
uniformly distributed around the sun: each class would be produced 
on one side of the sun or the other. Such clustering has been observed before as resulting
from the influence of NP [de la Fuente Marcos, 2017; Batygin, 2016b] 
as well as comets [Rickman et al., 2001]

In the previous encounters, the CWO stays far from the sun and its path is not affected much:
it will suffer a deviation of about a degree (or less). Its path is almost a straight line 
that is about perpendicular to 
the bisector of the angle between the two clusters (entrance and exit), 
both for EKO-I and EKO-II. 
Furthermore, as the orbits of the
EKO are stable, it is very difficult to estimate when this encounter took
place. From the positions of the objects discovered up to now on their respective 
orbit, it could be possible to estimate the number of cycles they have performed if 
we assume that they were produced at about the same time since the WCO crosses our solar 
system in a relatively short period.  

There is also a recurrent pattern to be noticed: 
for both EKO-I and EKO-II produced in this process, when $q$ decreases, 
$Q$ increases. This holds as long as the EKO-I and EKO-II produced originate from the 
same initial KO, the only difference being where the CWO interacts with the KO on its 
orbit. If we allow for the initial KO orbit to vary ($q,Q,\theta_x$), this relation 
will not quite hold when we compare the different EKO produced. If the initial parameters 
of the different KO are similar, the relation should hold relatively well.

\hfil\hfil\break
\noindent{\bf {Grazing collision}\hfil\hfil}
\hfil\hfil\break\noindent
We have studied a collision where the WCO cuts slightly the Kuiper belt.
As the collision becomes more and more grazing, (points A and B, or C and D get
closer) the window to produce EKO-II
decreases and the range in $q$ will also become very small. Eventually, when
the collision is almost tangential, it will be impossible to produce EKO-II 
while it will always be possible to produce EKO-I orbits. In that configuration, all 
objects sent towards the sun are ejected from the solar system, there is no stable EKO-II 
orbit.

\hfil\hfil\break
\noindent{\bf {Second Encounter}\hfil\hfil}
\hfil\hfil\break\noindent
Since the previous encounter cannot produce an orbit where $q> Q_{original}^{kuiper}$ and several  
of these obejcts have been observed (Class-III objects) we will allow the WCO to intersect 
the orbit of a typical EKO-I: 
$q_{EKO-I} = 6.4\times 10^{12}m$ and $Q_{EKO-I} = 5.67\times 10^{13}m$, which leads 
to $\varepsilon = 0.797$. We will let the distance from the sun ($x$, since the sun is at the 
origin) vary from $3\times 10^{13}$ to $5\times 10^{13}$; the $y$ coordinate being dictated by 
the original KO trajectory. The speed of the CWO will be $(0,50000)m/sec$, as before. The CWO is 
affected by the sun, but very little; its deflection being about 1 degree. For example, when 
it starts at $ x = -5\times 10^{13}$m, its closest distance to the sun will be about 
$-4.99\times 10^{13}$ m. Of course, the impact parameter between the CWO and the KO is much smaller.  
Figure 2 and Table 3 show orbits that such encounters can produce; the original orbit is the little 
ellipse at the center. The larger orbit is similar to 2015TG387, the second one is similar to 
2013SY99, the third one is similar to 2003VB12 (Sedna) and the fourth one is similar to 
2012VP113; these are the FKO-I. 
We also show another interesting orbit that is very 
large, has a relatively small 
eccentricity and runs clockwise; this is the FKO-II.. 
It is important to note that in order to produce 
2003VP113, the CWO came from the lower right such that its  
initial velocity was (-25.0,25.0) km/sec so that its velocity is about 35 km/sec 
as opposed to the usual 50 km/sec. This produced the desired orbit while staying far 
enough from the sun (at least 80 AU) so as to not disturb the Kuiper belt. Similarly, in 
order to produce the clockwise rotating orbit, the initial velociy of the CWO was 
$(-25.0, 40.0)km/sec$ which allowed it to stay far from the sun (at least 80 AU). 
One can see a pattern in the production of the FKO-I: when the interaction CWO-EKOI takes 
place closer to the sun, it will have a tendency to produce FKO-I with larger values of $q$. 
For similar values of $q$ there is a fair range of values for $Q$ by adjusting the collision 
parameters since all these encounters 
can eject the EKO-I when the CWO gets too close to the KO.

In order to produce the four FKO-I, a single WCO would have to cross the paths of 
four EKO-I. This scenario seems unlikely, but one can also take the point of view that if a 
first WCO produced several EKO-I, then a second object could cross their paths and produce 
something. In this view, the configuration that we have now is just one among several 
possibilities. The precise configuration of these 4 objects will depend where the CWO 
crossed the original EKO-I orbits and their orientation. The orbits that we present on Figure 2 
do not reproduce exactly those of these 4 objects because all three were produced from close 
points on the orbit of the original EKO-I. One can imagine however that these orbits can be 
reproduced in their configuration by allowing the CWO to cross the four EKO-I at different 
point on their respective orbit.

One must keep in mind also that the passage of a CWO in the neighbourhood of an FKO-I can 
realign its orbit by as much as 30-40 degrees, while leaving the other orbital parameters (q,Q) 
relatively unchanged. This way, a CWO passing relatively close to 2015TG387 when it is close to 
it aphelion could have modified 
its orientation while leaving Sedna and 2012VP113 unchanged; it might also have left 2013SY99 
unchanged, depending where it was on its orbit.

On Figure 3 and Table 4, we present some orbits that can be produced by the encounter of a WCO with 
an EKO-II. The EKO-II can remain in this class but its spin can be inverted. It can also 
become an EKO-I and again both spin directions are possible. It seems very difficult 
however to bring it back to a simple KO. Finally, it is possible to send it on very 
large orbits with relatively small eccentricities where both spinning directions are possible. 
These are the FKO-III class and they can be likened to the Lower Oort Cloud objects as their 
$Q$ becomes 
a fair fraction of a light year.

\hfil\hfil\break
\noindent{\bf {Two and three dimensions}\hfil\hfil}
\hfil\hfil\break\noindent
We have worked exclusively in two dimensions (2D) and one could argue that 
we have not reproduced all the orbits of the Sednitos discovered over the 
past 20 years; the most spectacular example being 2015BP519 which sticks out of 
the general orbital plane at 57 degrees or so. This poses a problem as 
Brasser et al. (2012) have shown that scattering interactions with the giant 
planets do not rise the orbital inclinations by much. As mentioned previously, 
this peculiar orbit could have been produced by NP but caution is 
required in this interpretation. However, as we have seen, the final orbital 
parameters of the perturbed objects depend greatly on the precise parameters 
of the encounter between the WCO and the initial object. No doubt that a 
slight difference in the encounter can have huge impact on 
the final parameters. For example, were the WCO to cross the orbital 
plane of the Kuiper belt instead of moving within it, this would have 
huge impacts and could very easily project the KO in an orbit 
perpendicular to the Kuiperian plane. Clearly, adding the third 
dimension complicates the problem substantially and we are confident that 
the main results provided in this paper would be valid in 3D. 
There is also an interesting mechanism discovered by Madigan 
[Madigan \& McCout, 2016, Madigan et al. 2018] called {\it inclination instability} 
where several eccentric orbits drive exponentially their inclination 
with respect to the solar plane. This process could potentially explain the very large 
inclination of 2015BP519. 

\hfil\hfil\break
\noindent{\bf {Mass, speed and zone of influence}\hfil\hfil}
\hfil\hfil\break\noindent
In this work, we have set the mass of the  messenger at 0.1 solar mass. 
In this way, it is barely a star and could have eluded detection up to now. 
We have verified that the zone of influence of this object on the Kuiper Belt 
is about 10 AUs in the sense that when the distance between the KO and 
the CWO is more than 10 AUs, the orbital parameters of the KO 
($q$ and $Q$) will remain within 10\% of their initial values, which 
means that it will remain as a standard KO. This definition is not as constraining as 
that of Gladman et al. (2008) regarding scattering objects but it seems 
reasonable for our purposes. If the KO  
object has a slightly eccentric orbit, the axes of the ellipse can be rotated 
by up to 90 degrees in some collisions, but 
the parameters will remain within 10\% of their initial values. Clearly, by reducing
the mass of the CWO, that radius of the zone fo influence will 
diminish, but so will the cross-section of interaction with the objects: the CWO 
will have to get closer to the KO in order to eject it 
out of the solar system or promote it to higher orbits.

The speed of the KO could also have an impact on the possibility 
of ejecting a KO out of the solar system or promoting it to 
higher orbits. This is another relatively free parameter that makes the prediction 
of the location of the CWO very difficult; we can estimate its direction from the 
clusters its passage has produced.

We have verified that the angle between the clusters (EKO-I or EKO-II) gets smaller 
(they become less back-to-back) as the collision between the CWO and the KO is moved 
farther away from the sun; it is about 140 degrees when the encounter takes place at 
about 80 AU, as opposed to about 170 degrees at 55 AU. 

\hfil\hfil\break
\noindent{\bf {Scenario}\hfil\hfil}
\hfil\hfil\break\noindent
When we take into account the following:
\hfil\hfil\break\noindent
\begin{itemize}
\item Scholz star has been observed and there might be more such objects
\item several EKO-II (or Class-II) objects have been 
discovered so far and their window of production is small and vanishes as the 
collision WCO-KO becomes tangential
\item several FKO-I (or Class-III) objects have been discovered so far 
and the production of a single FKO-I requires the interaction of 
a CWO with an EKO-I            
\item the production of several FKO-I through the passage of a single 
CWO seems unlikely and the passages of several WCO that would each produce a 
FKO-I seems also unlikely
\item clustering has been observed among the ETNO and clustering 
occurs naturally through the passage of a WCO across the out-skirt of the Kuiper Belt
\item the angle between clusters decreases as the point of interaction 
between the CWO and the KO is moved farther away from the sun (80 AU for example)
\item the FKO-I (or Class-III) objects observed up to now respect relatively 
well the relation observed here that  
when $q$ decreases, 
$Q$ increases 
\item the range of influence of a 0.1 solar mass CWO is about 10 AU 
\end{itemize}
\hfil\hfil\break\noindent
we are lead to consider the possibility that the Kuiper Belt might have been 
substantially larger than what it is now. In this scenario, a CWO would have cut through 
the out-skirt of this extended Kuiper belt. This would have depleted the belt 
of several members that would have been ejected but it would have also produced several 
objects in the EKO-I and EKO-II classes. The FKO-I (or Class-III) objects that 
are so difficult to produce from the actual Kuiper belt, would simply be EKO-I 
of the encounter ${\rm CWO-KO^{extended}}$. There could be one or two of these encounters, 
potentially over very long periods of time since the orbits are stable, on the 
same side of the sun or on opposite sides of the sun. Such a process likely would also create 
objects in classes that have not been observed yet; counter-rotating objects with 
very large semi-major axis and relatively small eccentricity for example. 
The resulting Kuiper belt would not 
be symmetric anymore and the asymmetrical bulges (sections of the belt that were left 
more or less untouched by the passages of the CWO) could be brought back in line with the 
belt by the gravitational forces of the belt over a long period of time. 
This scenario offers the beginning of a process where the Kuiper belt would eventually build a cut-off 
at about 50 AU.  

Interestingly, the {\it inclination instability} among eccentric orbits discovered by Madigan 
requires a large population of small objects beyond 50 AU and a few thousands AU with a 
total mass of 1-10 Earth masses to be able to operate. If this scenario is correct, the 
passage of a WCO at about 90 AU could have produced several exotic objects if the 
number of these objects is sufficiently large to make several encounters probable.

Clearly, if we want to consider an extended Kuiper Belt in this scenario, its width would 
have to be at least equal to the largest $q^{Class-III}$ which is about 80 AU. Furthermore, 
considering the EKO-II elements (mainly 2005VX3, 2012DR30, 2015ER61, and 2013BL76) one can 
see that there isn't half a plane that is not 
covered, which would indicate that there has been at least two WCO encounters, at least 
one on each side 
of the sun. Depending on how one combines these EKO-II into pairs, several orientation of 
the CWO would be possible. The clusterings might not be all that 
clear due to the different initial orbital parameters ($q,Q,\theta_x$) of the different KO affected 
in the process.  
This might also explain why the four Class-III objects listed in Table 1 do not respect the relation 
that when $q$ decreases, $Q$ increases: three elements respect this relation but a fourth one does 
not. This could mean that three elements had relatively close orbital paramters while the 
fourth one did not.


Consider now the clusterings presented on Figure 4 of Sheppard \& Trujillo (2019) 
(figure 1 of Batygin\& Morbidelli (2019) is similar) to which we add 
2015ER61, 2017DR30, 2015VX3 and 2013BL76. The main clustering 
presents two lobes consistent with what we have presented here; these would be EKO-I of an 
encounter ${\rm CWO-KO^{extended}}$ 
where the trajectory of the CWO would have been about perpendicular to the major axis of the 
Distant Eccentric Planet indicated on Figure 4. However, when we add 2015ER61 and 2017DR30 
we run into a problem because the 
EKO-II are more back-to-back than the EKO-I, which is contrary to what we have seen here. 
Another possibility is to consider one lobe aligned roughly with 2012VP113, another lobe 
aligned with the secondary cluster together with 2017DR30 and 2015VX3. This pattern could arise 
from the passage of a CWO with a trajectory at about 45 degrees with the major axis of the 
Distant Eccentric Planet. A second CWO parallel to the first one could explain 2015ER61, 
2013BL76 and a lobe aligned roughly with Sedna but there would be a lobe missing between 
the Secondary cluster and 2017DR30. Of course, if the CWO does not encounter any KO susceptible 
to produce an EKO-I, there could be a lobe missing.  
Clearly, there are several possibilities and
a full simulation involving hundreds or thousands of KO with a 
random angular
distribution but a tapering-off radial distribution extending up to 90-100 AU and being the
target of a CWO that would slice through it at about 85 AU would be very interesting !

Lastly, it is worth mentioning that it is possible to produce FKO-III , similar to those one 
could expect from a lower Oort cloud object (LOCO) not by interacting 
with the FKO-II itself, but with the sun. 
This is essentially conservation of angular momentum; it is shown on Figure 4 and Table 4 where the original path of the FKO-II is the small ellipse at the center; it spins counter-clockwise. 
The CWO is closest to the sun when the FKO-II is at the point where its orbit changes; close to its aphelion.  
The path where the CWO is deflected clockwise with respect to its original direction  
produces the FKO that rotates clockwise (red orbit) since the sun will acquire a counter-clockwise 
angular momentum in this collision. Conversely, the path where the Cosmic 
Messenger is deflected counter-clockwise with respect to its original direction produces the FKO that 
rotates counter-clockwise (green orbit), the same rotation as the original orbit, but with a 
much larger aphelion. 
Note that in both cases, the FKO is 
very far from the CWO and the sun when these two collide and the minimal distance between the CWO and the FKO-II is comparable to that between the sun and the FKO-II at the moment of the collision. Clearly, such an encounter would wreak havoc within the 
inner solar system and such effects would be seen on the motion of the inner planets. 

\hfil\hfil\break
\noindent{\bf {Conclusions}\hfil\hfil}
\hfil\hfil\break\noindent
We have studied the effects of the passage of a light WCO  
intersecting the outer edge of our Kuiper belt. We have set the mass 
of the WCO at 10\% of a solar mass 
so that it is barely a star and practically invisible.

We have seen that a WCO that crosses the orbit of a KO can eject it from the solar system,
but it can also send it on two different types of orbits: EKO-I that have perihelions similar 
to but smaller than $Q_{initial}^{Kuiper}$ and EKO-II whose perihelion is much smaller and can 
be just a few AU. EKO-I have a wider range of eccentricities while EKO-II tend to have very 
large eccentricities. Both will be produced in clusters and will obey the rule 
that if $q$ decreases, $Q$ increases. It is impossible to produce an orbit such that 
$q_{new} > Q_{initial}^{Kuiper}$. We have seen that objects like Sedna can be produced through 
the interaction of a CWO with an EKO-I object. However, considering the fact that 
each Sednito requires the interaction of a CWO with an EKO-I, it seems very unlikely 
that several CWO would have produced the Sednitos that we have observed up to now. Instead, it 
seems more appealing to consider the scenario where the Kuiper Belt was larger than what it is 
now, maybe not uniformed either so that current Sednitos are in fact EKO-I of an encounter 
between a CWO and an extended Kuiper belt. 
Most of the Sednitos observed up to now respect the rule that when 
$q$ increases, $Q$ decreases, indicating that the original KO orbits were fairly 
close to each other. The process studied here 
has a tendency to produce objects in clusters and the angle between the clusters decreases 
as the point of interaction CWO-KO moves farther away from the sun.
These observations point to an interaction that took place farther than 50 AU. 
This would require the extended Kuiper belt 
to reach about 90-100 AU and the encounter between the CWO and the KO to take place at about 85 AU. 
Furthermore, considering the current EKO-II population where much 
less than half of the plane is unoccupied, this mechanism would require at least two WCO, 
at least one on each side of the sun, to interact with the out-skirt of the extended Kuiper Belt. 
The scenario where a CWO skims an extended Kuiper Belt offers also the beginning of a mechanism that would 
eventually produce a sharp edge to the remaining Kuiper Belt.


The same way that the passage of Scholz's star seems to have left some traces in the radiants that have 
been observed, it is likely that the passage of a few CWO would also leave traces in the radiants, even though the mass that we used here for our CWO is less than that of Scholz's system.

\hfil\hfil\break\noindent
\hfil\hfil\break
\noindent{\bf {Acknowledgements}\hfil\hfil}
\hfil\hfil\break\noindent
I want to thank my colleague C.  Hamzaoui for interesting, 
enjoyable and stimulating discussions and G. Hellou for comments on the manuscript. 
I also want to thank my colleagues from the Atlas Collaboration 
at the physics department at Universit\'e de Montr\'eal for the use of their 
computers.  
\hfil\hfil\break
\hfil\hfil\break
\noindent{\bf {References}\hfil\hfil}
\hfil\hfil\break\noindent
Allen,R.L., Bernstein,G.M., Malhotra, R., {\it ApJ. 549},L241-L-244 (2001)
\hfil\hfil\break\noindent
Bacci,P.,Maestripieri,M.,Tesi,L. {\it et al.},2017,MPEC 2017-U181
\hfil\hfil\break\noindent
Bannister,M.T.,Shankman,C.,Volk,K.,{\it et al.},2017.arXiv1704.01952v1[astro-ph.EP]
Bannister,M.T.,Schwamb,M.,Fraser,W.C.,{\it et al.},2017b, {\it ApJ,851},L38
\hfil\hfil\break\noindent
Bannister,M.T.,Alexandersen,M.,Benecchi,S.D. {\it et al.} 2016,{\bf AJ},152:212
\hfil\hfil\break\noindent
Batygin,K., Brown, M.E.;2016,{\it AJ,151},22
\hfil\hfil\break\noindent
Batygin,K., Brown, M.E.; 2016b,{\it ApJ, 833}, L3
\hfil\hfil\break\noindent
Batygin,K.,Morbidelli,A., 2017, arXiv:1710.01804v1[astro-ph.EP]
\hfil\hfil\break\noindent
Becker,J.C., Khain,T., Hamilton,S.J.,{\it et al.},2018, {\it AJ},156,2
\hfil\hfil\break\noindent
Brasser, R.,Schwamb,M.E.,Lykawka,P.S., \& Gomes,R.S.
\hfil\hfil\break\noindent\phantom{sssss}
2012, {\it MNRAS}, 420, 3396
\hfil\hfil\break\noindent
Brasser,R,Schwamb,M.,{\it MNRAS}, 446, 3788, 2015
\hfil\hfil\break\noindent
Brasser,R.,Morbidelli, A.,2013, {\it Icarus}, 225,pp 40-49
\hfil\hfil\break\noindent
Brasser,R.,Duncan,M.J.,Levison,H.F.,Schwamb,M.E.,Brown,M.E.,2012,
\hfil\hfil\break\noindent\phantom{sssss}{\it Icarus},217, 1-19
\hfil\hfil\break\noindent
Bromley,B.C., Kenyon,S.J., 2016,{\it ApJ,826},64
\hfil\hfil\break\noindent
Brown,M.E., Trujillo, C., Rabinowitz,D., 2004; {\it ApJ,617}, 589-599
\hfil\hfil\break\noindent
Fernandez,J.A.,Brunini,A., {\it Icarus 106},580-590 (2000)
\hfil\hfil\break\noindent
Fraser,W.C.,Bannister, M.T.,Pike, R.E. {\it et als},2017,
\hfil\hfil\break\noindent\phantom{sssss}{\it Nature-Astronomy, 1},0088
\hfil\hfil\break\noindent
de la Fuente Marcos, C.,R. de la Fuente Marcos,Aarseth, S.J., 2018;
\hfil\hfil\break\noindent
\phantom{sssss}{\it MNRAS,476},(1),L1-L5
\hfil\hfil\break\noindent
de la Fuente Marcos, C.,R. de la Fuente Marcos,2018b,{\it RNAAS,2} 2018
\hfil\hfil\break\noindent
de la Fuente Marcos, C.,R. de la Fuente Marcos, 2017;
\hfil\hfil\break\noindent
\phantom{sssss}{\it MNRAS,471},(1),L61-L65
\hfil\hfil\break\noindent
de la Fuenta Marcos, C., de la Fuenta Marcos, R.,2014, {\it MNRAS,443}, L59
\hfil\hfil\break\noindent
Fienga, A.,Laskar,J.,Manche,H.,Gastineau,M., 2016;{\it A\&A,587},L8
\hfil\hfil\break\noindent
Gladman,B.,Kavelaars,J.J.,Petit,J.-M.,Morbidelli, A.,Holman,M.J.,Loredo,T.,2001,
\phantom{sssss}{\it AJ,122},1051-1066
\hfil\hfil\break\noindent
Gladman,B.,Holman,M.,Grav,T. {\it et al},2002, {\it Icarus,157}269-279
\hfil\hfil\break\noindent
Gladman, B., Marsden, B. G.,Vanlaerhoven, C.,
\hfil\hfil\break\noindent\phantom{sssss}{\it Nomenclature in the Outer Solar System},
 ed.M.A.Barucci, H. Boehnhardt,
 \hfil\hfil\break\noindent\phantom{sssss}
D. P. Cruikshank, A. Morbidelli, Dotson, R., 43-57,2008
\hfil\hfil\break\noindent
Gomes, R.S. 2003, {\it Icarus, 161}, 404-418
\hfil\hfil\break
Gomes, R., Fernandez, J., Gallardo, T., Mrumini, A., 2008, 
\hfil\hfil\break\noindent\phantom{sssss}{\it The Solar System beyond Neptune}
\hfil\hfil\break\noindent\phantom{sssss}
ed. M. Barucci et al., (Tucson, AZ., Univ. Arizona Press), pp 259-273
\hfil\hfil\break\noindent
Hands,T.O.,Dehnen,W.,Gration,A.,Stadel,J.,Moore,B.,2019, {\it MNRAS} april
\hfil\hfil\break\noindent
Holman,M.J.,Payne,M.J.,Fraser,W.,{\it et al.},2018,{\it ApJ,855},L6
\hfil\hfil\break\noindent
Ida,S.,Larwood,J.,Burkert,A.,2000,{\it ApJ,528},351-356
\hfil\hfil\break\noindent
Jilkova,L.,Portegies Zwart,S.,Tjibaria,P.,Hammer,M.;2015,
{\it MNRAS,453},(3),3157
Kaib,N.A.,Pike,R.,Lawler,S.,{\it el al.},2019,arXiv:1905-09286v2[astro-ph.EP]
\hfil\hfil\break\noindent
Kenyon,S.J.,Bromly,B.C.,2004, {\it Nature,432}, 598-602
\hfil\hfil\break\noindent
Knight, M.M.,Protopapa, S., Kelley, S.S.P., et al., 2017,{\it ApJ,851}, L31
\hfil\hfil\break\noindent 
Lawler,S.M.,Shankman,C.,Kaib,N., {\it et al.},2016,{\it AJ,153}, 1
\hfil\hfil\break\noindent
Levison, H.F., Morbidelli,A.,van Laerhoven, C.,Gomes,R.,2007,
\hfil\hfil\break\noindent\phantom{sssss}{\it Icarus,196}, 258-273
\hfil\hfil\break\noindent
Li,G.,Adams,F.C.,2016,{\it ApJ,823},L3.
\hfil\hfil\break\noindent
Linder,E.F.,Mordasini,C.,2016;{\it A\&A,589},A134
\hfil\hfil\break\noindent
Madigan, A.-M., McCourt, M.,2016,{\it MNRAS,457}, L89
\hfil\hfil\break\noindent
Madigan,A.-M., Zderic, A.,McCourt,M., Fleisig, J.,2018, {\it AJ,156},4
\hfil\hfil\break\noindent
Malhotra, R.,Volk,K.,Wang,X.,2016;{\it A\&A,589},A134
\hfil\hfil\break\noindent
Mamajek,E.E.,Barenfeld,S.S.,Ivanov,V.D.,Kniazev,A.Y.,Vaisanen,P.,
\hfil\hfil\break\noindent\phantom{ssss}
Beletsky,Y.,Boffin,H.M.J.,2015,{\it ApJ,800},L17
\hfil\hfil\break\noindent
Meech,K.,Bacci,P.,Maestripieri,M.,{\it et al.},2017,{\it MPEC 2017-U183.A/2017} U1
\hfil\hfil\break
Meech, K.J., et al. 2017b, {\it Nature,552}, 378 
\hfil\hfil\break\noindent
Millholland,S.,Laughlin,G.,2017,{\it AJ,153},91
\hfil\hfil\break\noindent
Morbidelli, A., Emel'yanenko,V., Levison, H.F.,2004, {\it MNRAS,355},935-940
\hfil\hfil\break
Morbidelli, A.,Nesvorny,D.; 2019 arXiv:1904.02980v1[astro-ph.EP]
\hfil\hfil\break\noindent
Morbidelli, A.,Levison,H.F.,2004,{\it ApJ,128},2564
\hfil\hfil\break\noindent
Morbidelli, A.,Levison, H.F., Gomes, R, 2007 arXiv:astro-ph/07-3558v1
\hfil\hfil\break\noindent
Pfalzner,S.,Bannister,M.T.,2019, {\it ApJ, 874},L34
\hfil\hfil\break\noindent
Rickman, H.,Valsecchi, G.B., Froeschlé, C., 2001,{\it MNRAS,325}, 1303
\hfil\hfil\break\noindent 
Scholz,R.-D.,2014,{\it A\&A,561},A113
\hfil\hfil\break\noindent
Shankman,C.,Kavelaars,J.J.,Bannister,M.T.,Gladman,B.J.,Lawler,S.M.,
\hfil\hfil\break\noindent\phantom{ssss}
Chen,Y.-T.,Jakubik,M.,Kaib,N.,Alexandersen,M.,Gwyn,S.D.J.,
\hfil\hfil\break\noindent\phantom{sssss}
Petit,J.-M., Volk,K.,2017, {\it AJ 154}, 50
\hfil\hfil\break\noindent
Sheppard,S.S.,Trujillo, C., Tholen,D.J,2016, {\it ApJ,825}, L13
\hfil\hfil\break\noindent
Sheppard,S.S.,Trujillo,C.A.,2016b, {\it AJ,152}, 221
\hfil\hfil\break\noindent
Sheppard, S.S.,Trujillo,C.A.,Tholen,D.J.,Kaib,N.,2019, {\it AJ,157},139
\hfil\hfil\break\noindent
Singer,K.N.,McKinnon,W.B.,Gladman,B.,{\it et al.}, 2019, {\it Science,363},955
\hfil\hfil\break\noindent
Stern,S.A.,2004, arXiv:astro-ph/0404525v2
\hfil\hfil\break\noindent
Thirouin,A.,Sheppard,S.S.,2018;arXiv:1804.09695v1[astro-ph.EP]
\hfil\hfil\break\noindent
Thirouin,A.,Sheppard,S.S.,Noll,K.S.,2017;{\it ApJ,844},135
\hfil\hfil\break\noindent
Trujillo, C.A., Brown, M.E., 2001,{\it AJ, 554}, L95-L98
\hfil\hfil\break\noindent
Trujillo,C.A.,Sheppard,S.S., 2014,{\it Nature,507},471
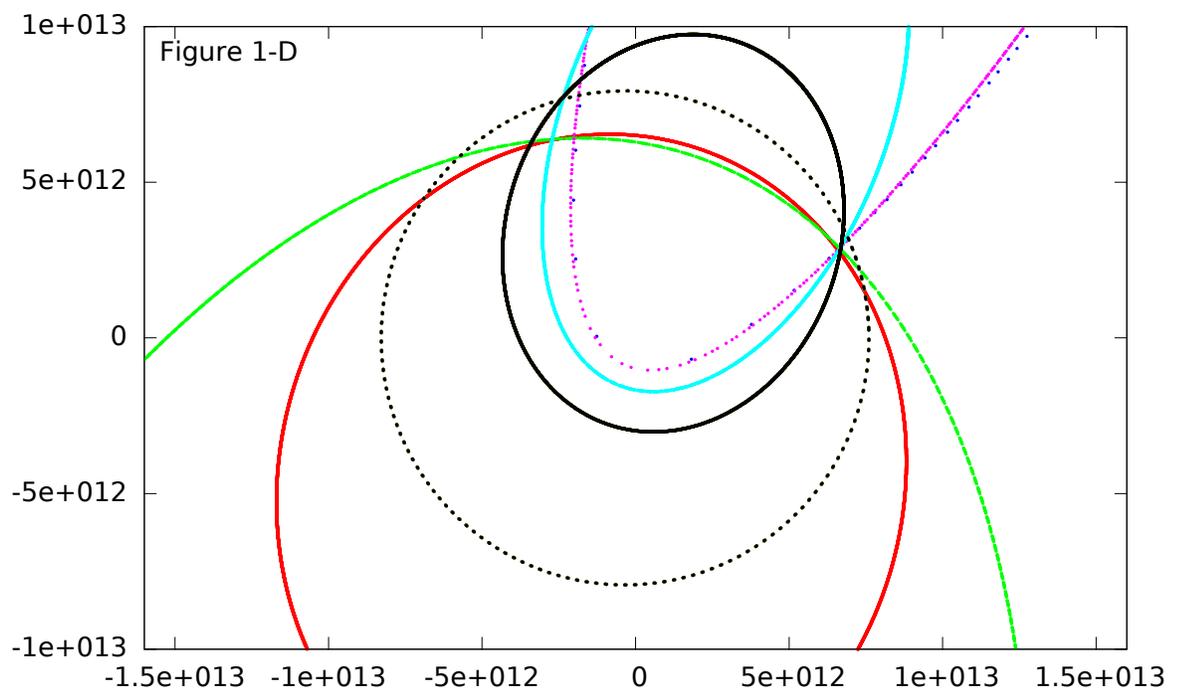
\hfil\hfil\break\noindent

\hfil\hfil\eject

\begin{figure}[t]
\includegraphics[scale=1.25]{figure1A}
\noindent
Orbits produced at the entrance in a head-on collision 
between a CWO and a KO.
\end{figure}
\vfil\vfil\eject

\begin{figure}[t]
\includegraphics[scale=1.25]{figure1B}
\noindent
Orbits produced at the exit in a head-on collision
between a CWO and a KO.
\end{figure}
\vfil\vfil\eject

\begin{figure}[t]
\includegraphics[scale=1.25]{figure1C}
\noindent
Orbits produced at the entrance of a rear-end collision 
between a CWO and a KO.
\end{figure}
\vfil\vfil\eject

\begin{figure}[t]
\includegraphics[scale=1.25]{figure1D}
\noindent
Orbits produced at the exit of a read-end collision 
between a CWO and a KO.
\end{figure}
\vfil\vfil\eject

\begin{figure}[t]
\includegraphics[scale=1.2]{figure2}
\noindent
Orbits produced in the collision between a CWO and an EKO-I 
that reproduce 4 known exotic objets. 
\end{figure}
\vfil\vfil\eject

\begin{figure}[t]
\includegraphics[scale=1.25]{figure3}
\noindent
Orbits produced in the collision between a CWO and an EKO-II.
\end{figure}
\vfil\vfil\eject

\begin{figure}[t]
\includegraphics[scale=1.25]{figure4}
\noindent
Orbits produced from an EKO-II in the collision between a CWO and the Sun.
\end{figure}
\hfil\hfil\eject
\begin{table}[h!]
 \begin{center}
 \caption{Orbital parameters of known objects}
  \begin{tabular}{c|c|c}
   \hline
   {\bf Element} &{\bf q(AU)} & {\bf Q(AU)}\\
   \hline
   \multicolumn{3}{c}{Class I}\\
   \hline
  2004VN112 & 47.3 & 585\\
  2014FZ71  & 56   & 97\\
  2013RF98  & 36   & 662\\
  2000OO67  & 20   & 1068\\
  2006SQ372 & 24.1 & 1461\\
  2013FY27  & 37   & 82\\
  2015BP519 & 35   &  824\\
  2015KG163 & 40   & 1581\\
  2015GT50  & 38   & 631\\
  2014FE72  & 36   & 3850\\
  2014SR349 & 47   & 535\\
\hline
\multicolumn{3}{c}{Class II}\\
\hline
   2005VX3 & 4.13   & 3079\\
   2002RN109 &2.7  & 1155\\
   2013BL76 & 8.36 & 2151\\
   2010BK118& 6.1  & 963\\
   2010NV1  & 9.3  & 596\\
   2015ER61 & 1.06 & 2487\\
   2007DA61 & 2.67 & 990\\
   2013AZ60 & 7.92 & 1117\\
   2012DR30 & 14.6 & 2939\\
\hline
\multicolumn{3}{c}{Class III}\\
\hline
   2003VB12  & 76  & 954\\
   2012VP113 & 80.5 & 445\\
   2015TG387 & 65   & 2275\\
   2013SY99  & 50   & 1420\\
   \hline
   \multicolumn{3}{l}{this list is not exhaustive}\\
   \multicolumn{3}{l}{but representative}\\
\end{tabular}
\end{center}
\end{table}

\vfil\vfil\eject

\begin{table}[h!]
 \begin{center}
 \caption{Orbital parameters}
  \begin{tabular}{c|c|c|c}
   \hline
   {\bf Element} &{$q({m\atop AU})$} & {$Q({m\atop AU})$} & $\theta_x^Q$\\
   \hline
   \multicolumn{4}{c}{Figure 1}\\
    \hline 
    initial  & $7.60\times 10^{12}\atop 50$ & $ 8.28\times 10^{12}\atop 55.2$ & 180\\
    \hline
    \multicolumn{4}{c}{Figure 1-A}\\
    \hline
    red     & ${2.18\times 10^{12}\atop 14.5}$  & ${1.35\times 10^{13}\atop 89.3}$ & -115\\
    green   & ${5.79\times 10^{11}\atop 3.9}$   & ${1.91\times 10^{15}\atop 12733}$& -127\\
    blue    & ${5.72\times 10^{12}\atop 38.1}$  & ${1.36\times 10^{15}\atop 9067}$ & +84\\
    magenta &${5.96\times 10^{12}\atop 39.7}$  & ${5.20\times 10^{13}\atop 347}$  & +85\\
    light blue &${6.26\times 10^{12}\atop 41.7}$ & ${2.21\times 10^{13}\atop 147}$ & +85\\
    black   &${6.57\times 10^{12}\atop 43.8}$    & ${1.14\times 10^{13}\atop 76}$ & +87\\
    \hline
    \multicolumn{4}{c}{Figure 1-B}\\
    \hline
     red     &${2.95\times 10^{12}\atop 19.7}$ & ${1.45\times 10^{13}\atop 96.7}$  &  +109\\
     green   &${1.06\times 10^{12}\atop 7.1}$  & ${2.51\times 10^{14}\atop 1673}$  &  +120\\
     blue    &${6.13\times 10^{12}\atop 40.9}$ & ${1.93\times 10^{15}\atop 12867}$ &  -84\\
     magenta &${6.28\times 10^{12}\atop 41.9}$ & ${1.19\times 10^{14}\atop 793}$   &  -83\\
     light blue &${6.75\times 10^{12}\atop 45}$ & ${2.93\times 10^{13}\atop 195}$  &  -82\\
      black &${7.35\times 10^{12}\atop 49}$ & ${1.39\times 10^{13}\atop 92.7}$ & -88\\ 
      \hline
      \multicolumn{4}{c}{Figure 1-C}\\
      \hline    
      red   &${7.00\times 10^{12}\atop 46.7}$ & ${4.21\times 10^{13}\atop 281}$   & +115\\
      green &${1.34\times 10^{12}\atop 8.9}$  & ${8.01\times 10^{13}\atop 5340}$  & -70\\
      blue  &${1.53\times 10^{12}\atop 10.2}$ & ${6.44\times 10^{13}\atop 429}$   & -70\\
      magenta&${3.75\times 10^{12}\atop 25}$   & ${1.21\times 10^{13}\atop 80.7}$  & -78\\
      light blue&${5.52\times 10{12}\atop 36.8}$ & ${1.05\times 10^{13}\atop 70}$  & -90\\
      \hline
      \multicolumn{4}{c}{Figure 1-D}\\
      \hline
      red &${6.41\times 10^{12}\atop 42.7}$    & ${1.61\times 10^{13}\atop 107}$    & -109\\
      green &${6.16\times 10^{12}\atop 41.1}$  & ${6.74\times 10^{13}\atop 449}$    & -107\\
      blue &${8.88\times 10^{11}\atop 5.9}$    & ${8.39\times 10^{13}\atop 5593}$   & +64\\
      magenta&${9.49\times 10^{11}\atop 6.3}$  & ${1.36\times 10^{14}\atop 907}$    & +65\\
      light blue&${1.62\times 10^{12}\atop 10.8}$ &${1.68\times 10^{13}\atop 112}$  & +67\\
      black &${2.92\times 10^{12}\atop 19.2}$  &${1.01\times 10^{13}\atop 67.3}$    & +70\\
      \hline
   \end{tabular}
   \end{center}
   \end{table}
    \vfil\vfil\eject
    
    \begin{table}[h!]
    \begin{center}
    \caption{Orbital parameters}
    \begin{tabular}{c|c|c|c|c}
    \hline
    {\bf Element} & $q({m\atop AU})$ & $Q({m\atop AU})$ & $\pm$ & $x_{interaction}(m)$\\
    \hline
    \multicolumn{5}{c}{Figure 2}\\
    \hline
    initial & ${6.40\times 10^{12}\atop 42.7}$ & ${5.66\times 10^{12}\atop 377}$ & + & ****\\
    \hline
    red & ${1.05\times 10^{13}\atop 70}$ & ${3.86\times 10^{14}\atop 2576}$ & + & $-5.0\times 10^{13}$\\
    green & ${8.50\times 10^{12}\atop 57}$ & ${2.19\times 10^{14}\atop 1462}$ & + & $-5.5\times 10^{13}$\\
    blue & ${1.03\times 10^{13}\atop 69}$  & ${1.45\times 10^{14}\atop 968}$  & + & $-4.5\times 10^{13}$\\
    magenta&${1.17\times 10^{13}\atop 78}$  & ${7.74\times 10^{13}\atop 516}$  & + & $-3.0\times 10^{13}$\\
    light blue&${3.42\times 10^{13}\atop 228}$ & ${5.10\times 10^{13}\atop 340}$ & -& $-3.3\times 10^{13}$\\
    \hline
    \multicolumn{5}{l}{$+$ is counterclockwise rotation}\\
    \multicolumn{5}{l}{$-$ is clockwise rotation}\\
    \end{tabular}
    \end{center}
    \end{table}
    
    \begin{table}[h!]
    \begin{center}
    \caption{Orbital parameters}
    \begin{tabular}{c|c|c|c}
    \hline
    {\bf Element} & $q({m\atop AU})$ & $Q({m\atop AU})$ & $\pm$\\
    \hline
    \multicolumn{4}{c}{Figure 3}\\
    \hline
    initial & ${8.3\times 10^{11}\atop 5.53}$ & ${8.28\times 10^{14}\atop 5520}$ & $+$ \\
    \hline
    red & ${1.63\times 10^{14}}$  & ${1.28\times 10^{15}\atop 8533}$  & $+$\\
    green & ${3.26\times 10^{13}\atop 217}$  & ${1.11\times 10^{15}\atop 7400}$ & $-$ \\
    \hline
    \multicolumn{4}{c}{Figure 4}\\
    \hline
    initial & ${8.3\times 10^{11}\atop 5.53}$  & ${8.28\times 10^{14}\atop 5520}$  &  $+$\\
    \hline
    red & ${3.81\times 10^{14}\atop 2540}$  &  ${2.38\times 10^{15}\atop 15867}$ & $-$\\
    green & ${2.22\times 10^{14}\atop 1480}$ & ${2.92\times 10^{15}\atop 19467}$ & $+$\\
    \hline
    \multicolumn{4}{l}{$+$ is counterclockwise rotation}\\
    \multicolumn{4}{l}{$-$ is clockwise rotation}\\
    \end{tabular}
    \end{center}
    \end{table}

\end{document}